\documentclass{ws-procs975x65}

\begin{document}
\title{THE SIGNATURE OF TIDAL DISRUPTION PHENOMENA \\ IN THE VICINITY OF A BLACK HOLE
}
\author{U. KOSTI\' C$^*$ and A. \v CADE\v Z}
\address{University of Ljubljana, Faculty of Mathematics and Physics\\
Jadranska 19, Ljubljana, Slovenia\\
$^*$e-mail: uros.kostic@fmf.uni-lj.si\\
www.fmf.uni-lj.si/en/}
\author{M. CALVANI and C. GERMAN\`A}
\address{INAF - Astronomical Observatory of Padova\\
Vicolo Osservatorio 5, 35122 Padova, Italy}
\begin{abstract}
Tidal effects on clumps of material during random non-stationary accretion onto a black hole produce phenomena with distinct temporal characteristics in observed light-curves. During such non-stationary accretion events, the shape of the accreting object evolves in time, and observable quasi-periodic phenomena with variable quasi-periods are produced. A number of characteristic light-curves, obtained with numerical simulations, will be shown. Their relevance to observed phenomena will be briefly discussed.
\end{abstract}
\bodymatter
%
\section{Conditions for strong tides and the Galactic center}
Strong tidal effects are induced by the gravitational field of a black hole on objects that find themselves below their Roche radius, leading to tidal disruption and to the release of a large part of gravitational energy on short time-scales. The strength of the tidal force is described by the Roche penetration factor $\beta = r_{\rm R}/ r_{\rm p}$, that is by the ratio of the Roche radius to the periastron distance.

For the black hole at the Galactic center, the condition of strong tidal forces $\beta \geq 1$ are met in the top left shaded area of \fref{GC_BH}(a). Objects in this area are tidally deformed and capable of releasing significant tidal energy. See Ref.~\refcite{2009A&A...496..307K} for more details.
\def\figsubcap#1{\par\noindent\centering\footnotesize(#1)}
\begin{figure}[h]%
\begin{center}
  \parbox{5.0cm}{\epsfig{figure=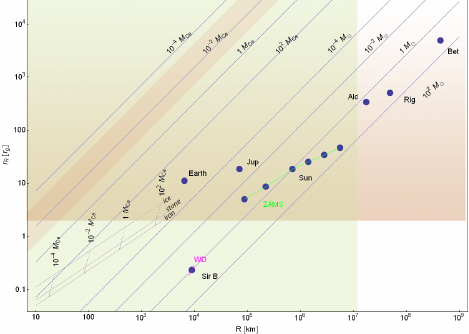,height=5.0cm}\figsubcap{a}}
  \hspace*{60pt}
  \parbox{5.0cm}{\epsfig{figure=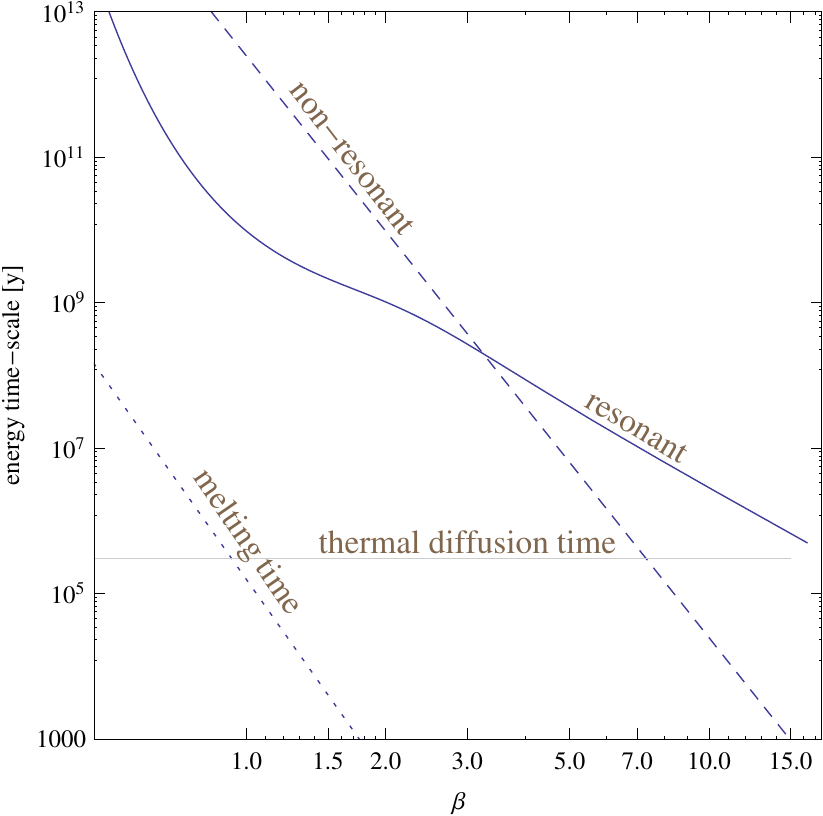,height=5.0cm}\figsubcap{b}}
  \caption{(a) Roche radius as function of mass and radius $R$ of the object. (b) Resonant and non-resonant time-scales as a function of $\beta$ ($\varrho = 1\ \mathrm{g cm^{-3} }$ and $R = 10\ \mathrm{km}$).}%
  \label{GC_BH}
\end{center}
\end{figure}
\section{Evolution of orbits and the free-fall phase}
Tidal torques on the infalling object are responsible for energy and angular momentum losses. Energy is dissipated on two different time-scales, resonant and non-resonant, that depend on $\beta$ as shown in \fref{GC_BH}(b). We have shown that low-mass satellites, like asteroids and comets, can get close to the black hole at the Galactic center. See Ref.~\refcite{2008A&A...487..527C} for more details.

We developed a relativistic numerical code to investigate the last stages of tidal evolution. The accreting body is initially treated as a rigid sphere outside $r_{\rm R}$ and as freely falling particles below $r_{\rm R}$. Figures~\ref{FF_above} and \ref{FF_equat} show how the luminosity of an object, infalling on a parabolic orbit, changes in time. 
Due to strong gravitational lensing, we included the signal from primary and secondary images in our calculations. The primary image is produced by photons taking the shortest path to the observer, and the secondary by photons passing ''behind'' the black hole before reaching the observer. Note also the strong effects of gravitational red-shift. Detailed videos are available from the authors. See Ref.~\refcite{2009A&A...496..307K} for more details.
\def\figsubcap#1{\par\noindent\centering\footnotesize(#1)}
\begin{figure}[h]%
\begin{center}
  \parbox{3.0cm}{\epsfig{figure=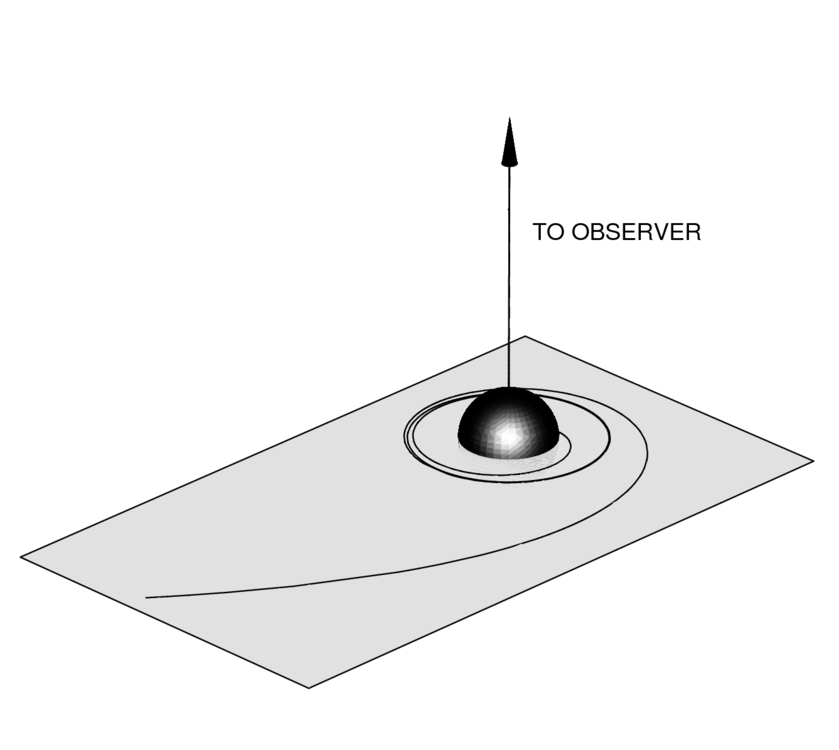,height=3.0cm}\figsubcap{a}}
  \hspace*{20pt}
  \parbox{3.5cm}{\epsfig{figure=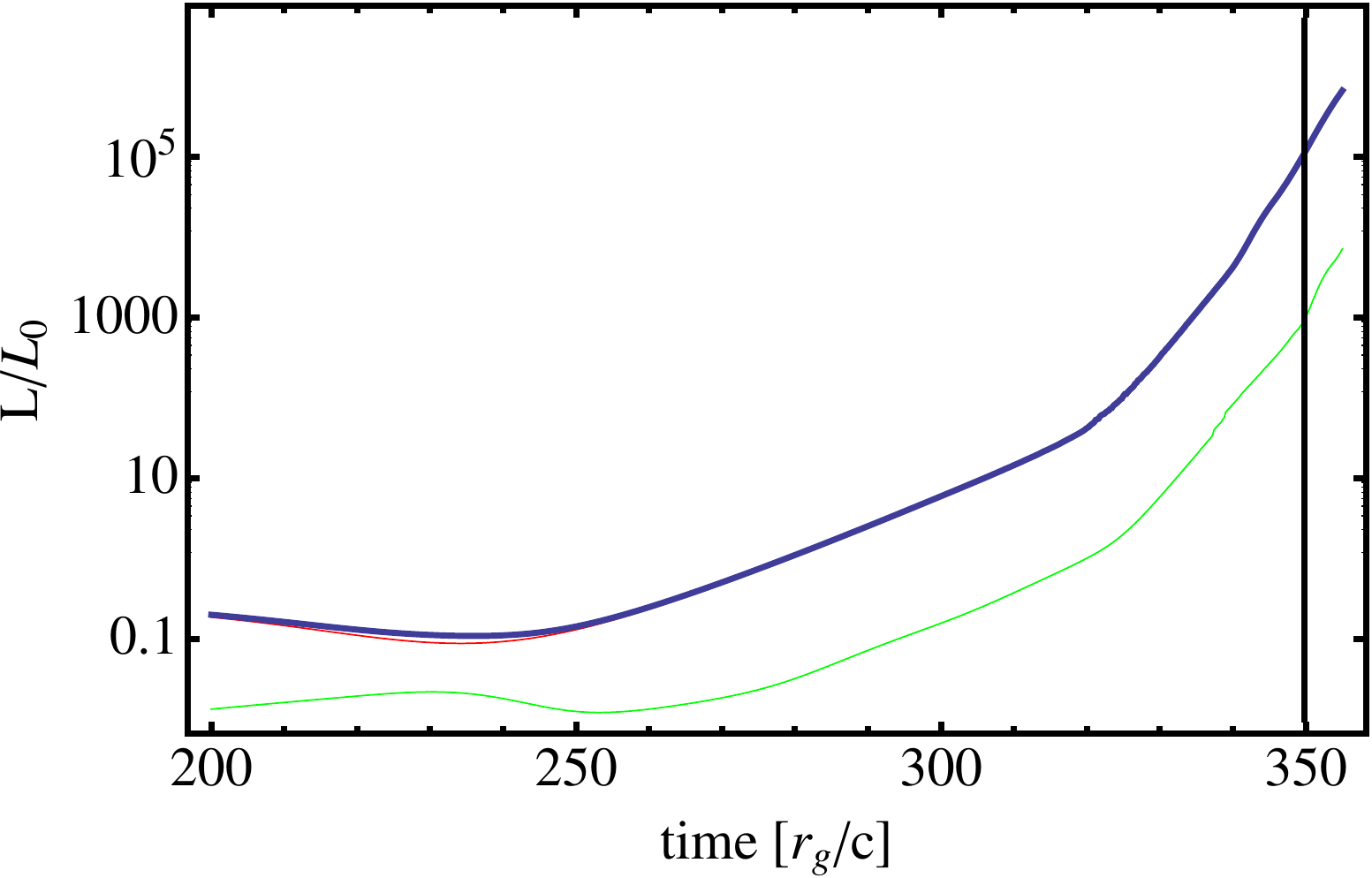,height=3.5cm}\figsubcap{b}}
  \caption{(a) The observer is exactly above the orbital plane. (b) Luminosity evolution: green line represents the signal from primary image, red line is from the secondary image, and blue is the sum of both signals.}%
  \label{FF_above}
\end{center}
\end{figure}
\begin{figure}[h]%
\begin{center}
  \parbox{3.0cm}{\epsfig{figure=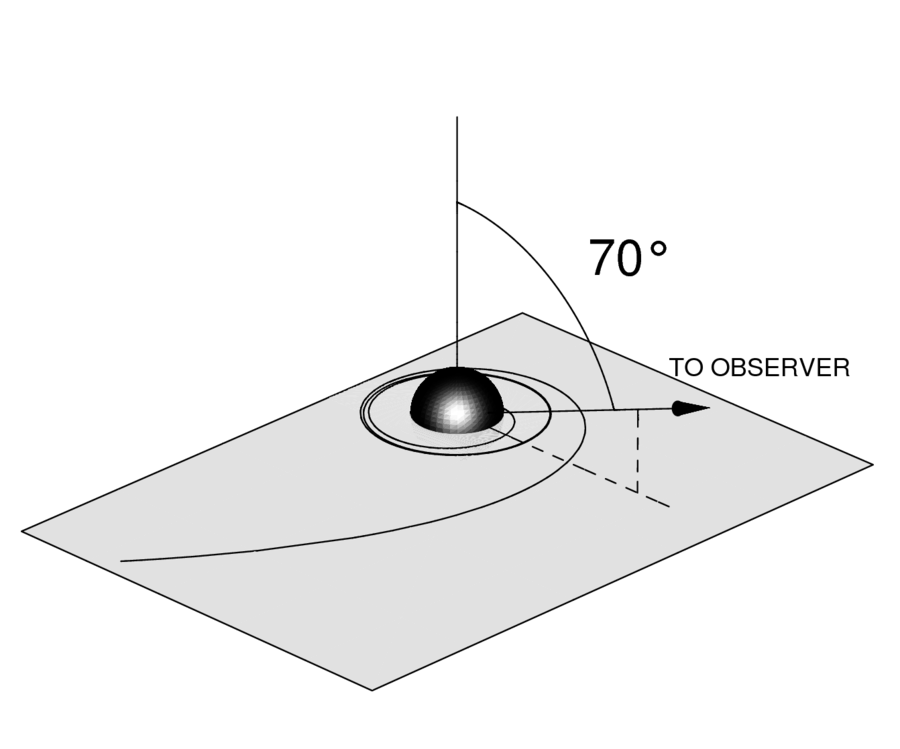,height=3.0cm}\figsubcap{a}}
  \hspace*{20pt}
  \parbox{3.5cm}{\epsfig{figure=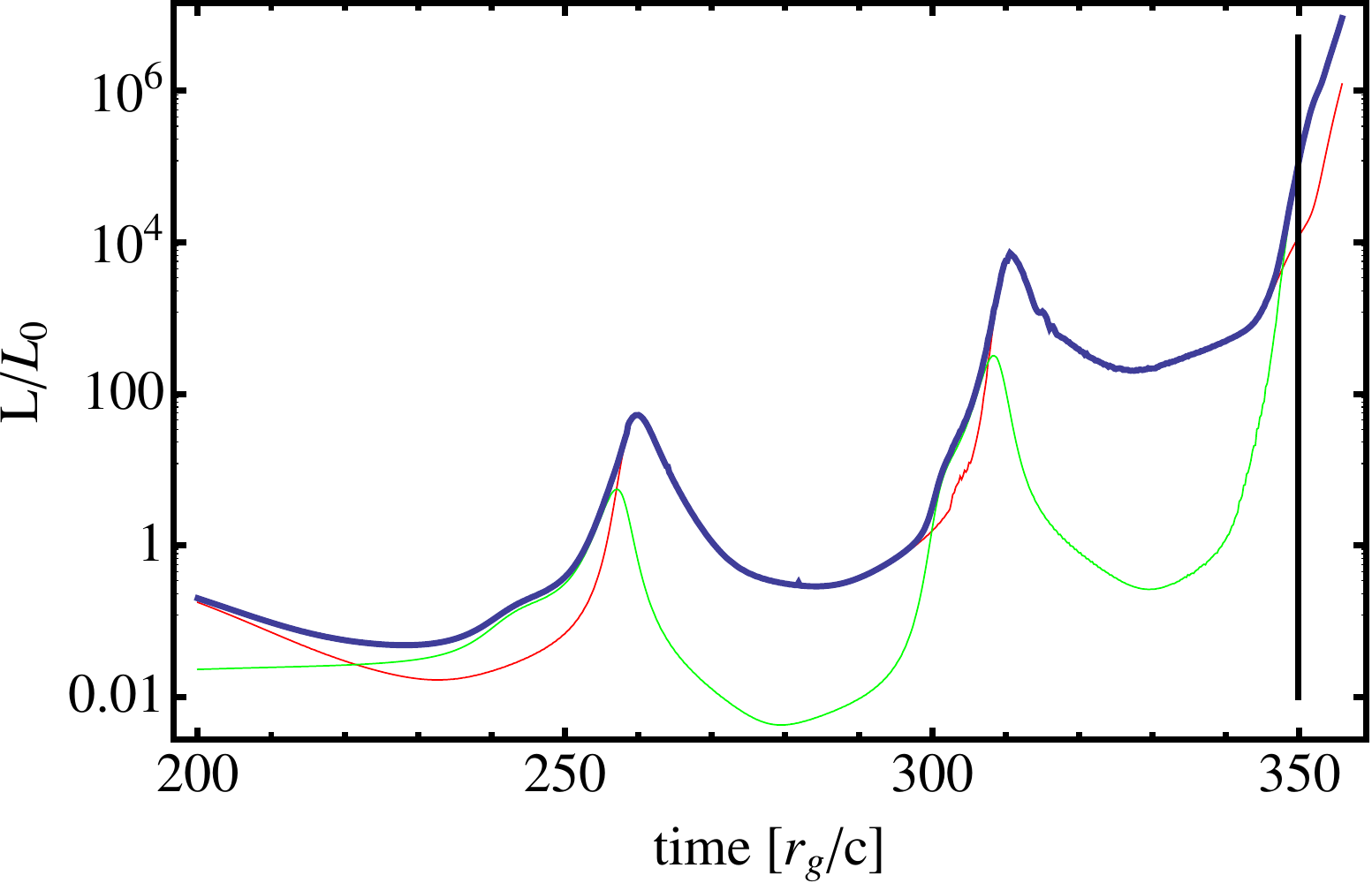,height=3.5cm}\figsubcap{b}}
  \caption{(a) The observer is $20^{\circ}$ above the orbital plane. (b) Luminosity evolution: green line represents the signal from primary image, red line is from the secondary image, and blue is the sum of both signals.}%
  \label{FF_equat}
\end{center}
\end{figure}
\section{Flares from the Galactic center and QPOs in X-ray binaries}
\begin{figure}[h]%
\begin{center}
  \parbox{5.5cm}{\epsfig{figure=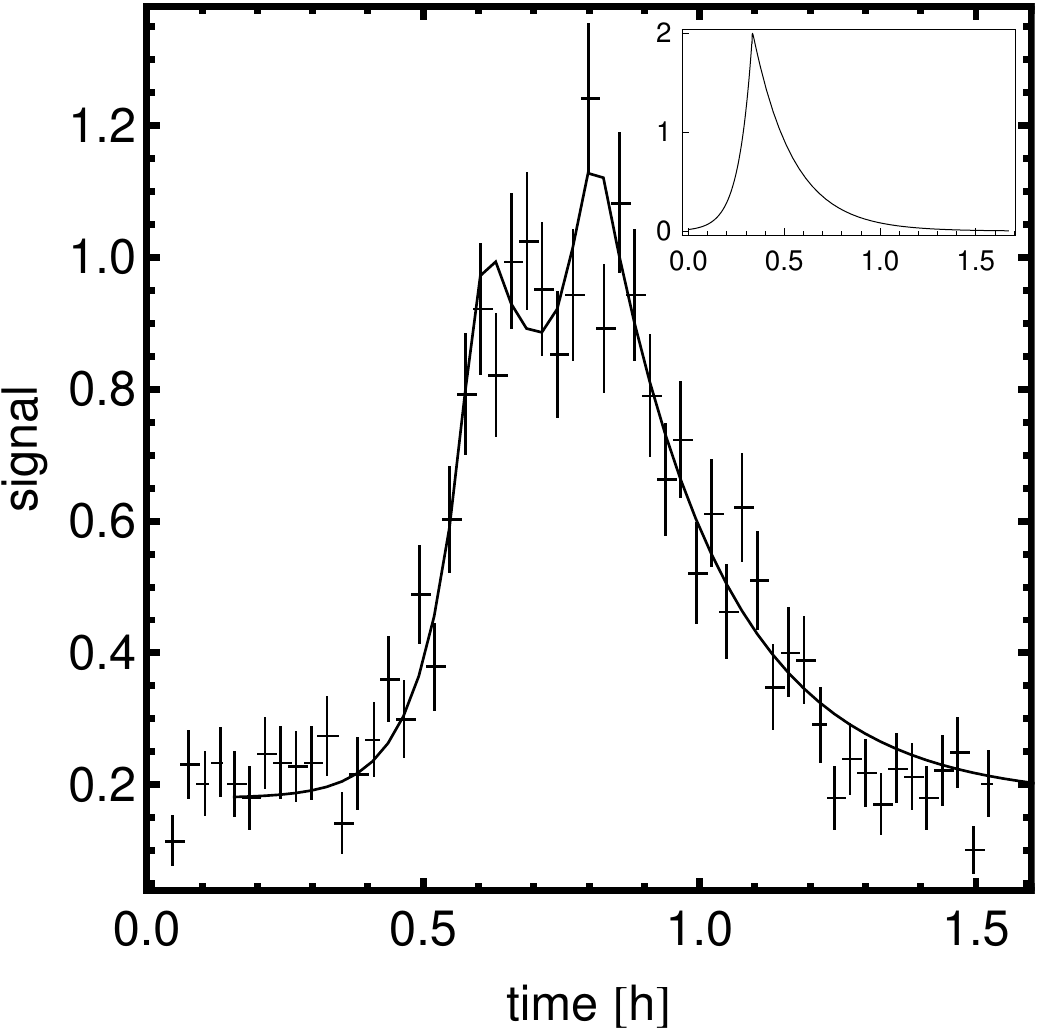,height=5.5cm}\figsubcap{a}}
  \hspace*{20pt}
  \parbox{5.5cm}{\epsfig{figure=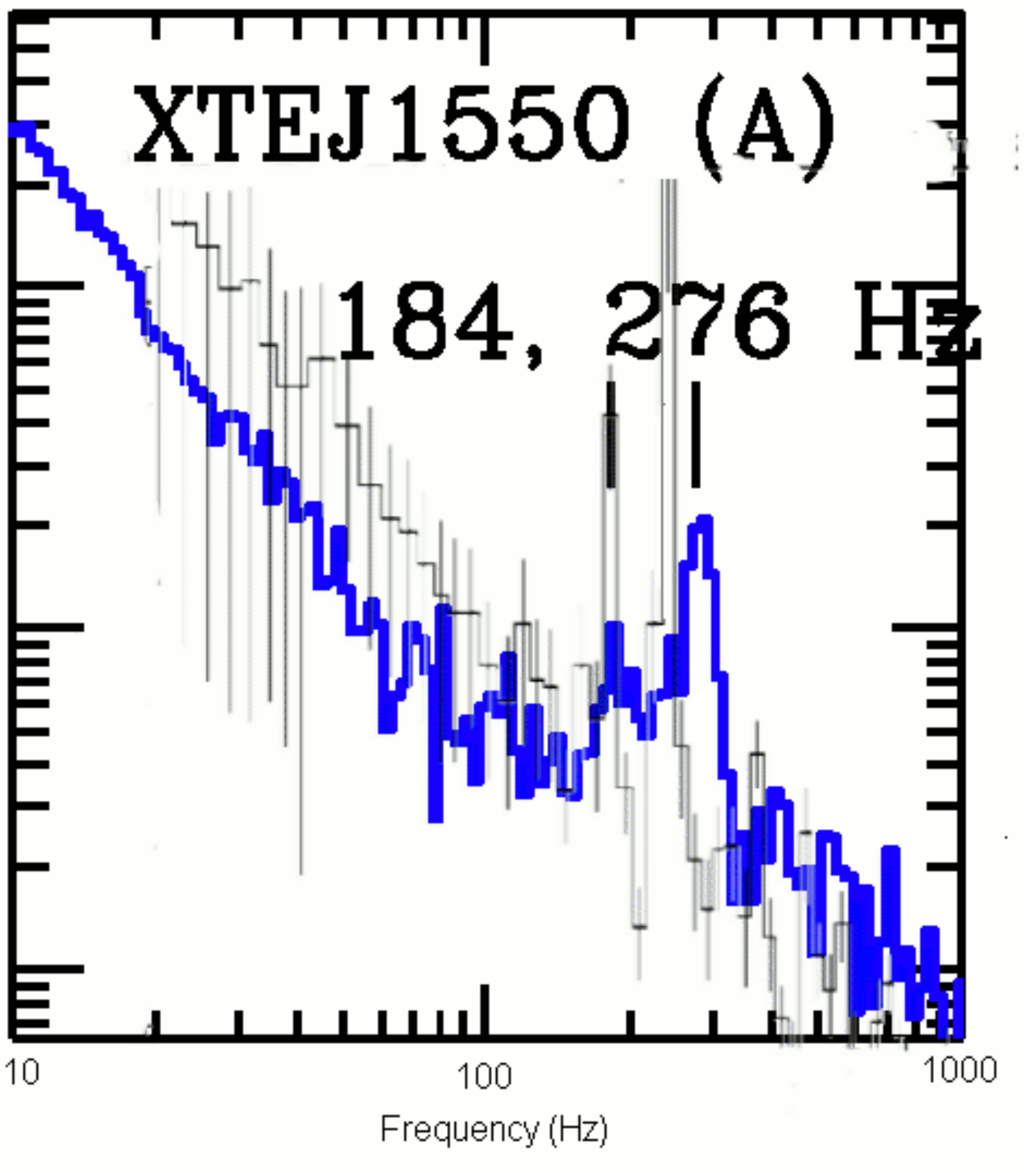,height=5.5cm}\figsubcap{b}}
  \caption{(a) The XMM-Newton/EPIC light curve of the April 4, 2007 flare and our simulated light-curve. (b) High frequency part of XTE J1550-564 power spectrum (blue) with our simulated power spectrum (black).}%
  \label{FLARE}
\end{center}
\end{figure}

The results of our numerical simulations were used as Green's function to calculate the light curve of flares from the Galactic center, see \fref{FLARE}(a). We also made a Fourier power spectrum of the signal to fit the high frequency part of the spectra observed in the LMXB XTE J1550-564 containing a black hole, see \fref{FLARE}(b). Furthermore, preliminary estimates on evolution of (internal) magnetic fields show that magnetic fields may change tidal deformations of the infalling material. Details in Refs.~\refcite{2009A&A...496..307K,2009AIPC.1126..367G}.

Summing up, one should consider not only accretion disks but also transient accretion phenomena to better explain XR/IR flares from the Galactic centre, QPOs, and any other non-stationary phenomena which may involve black holes.
%
\bibliographystyle{ws-procs975x65.bst}
\bibliography{/home/uros/biblio}
\end{document}